\documentclass[aps,showpacs,twocolumn,eqsecnum,superscriptaddress]{revtex4}

\usepackage{epsfig}
\usepackage{latexsym}

\begin{document}


\title{Testing excision techniques for dynamical 3D black hole
evolutions}


\author{Miguel Alcubierre}
\affiliation{Instituto de Ciencias Nucleares, Universidad
Nacional Aut\'onoma de M\'exico, A.P. 70-543, M\'exico D.F. 04510,
M\'exico.}

\author{Bernd Br\"ugmann}
\affiliation{Institute for Gravitational Physics and Geometry, Penn State
University, University Park, PA 16802, U.S.A.}

\author{Peter~Diener}
\affiliation{Department of Physics and Astronomy,
Louisiana State University, Baton Rouge, LA 70803, USA}
\affiliation{Center for Computation and Technology, 302 Johnston Hall,
Louisiana State University, Baton Rouge, LA 70803, USA}

\author{Frank~Herrmann}
\author{Denis~Pollney}
\affiliation{Max Planck Institut f\"ur Gravitationsphysik,
Albert Einstein Institut, Am Muehlenberg 1, 14476 Golm, Germany}

\author{Ed~Seidel}
\affiliation{Department of Physics and Astronomy, Louisiana State University,
Baton Rouge, LA 70803, USA}
\affiliation{Center for Computation and Technology, 302 Johnston Hall,
Louisiana State University, Baton Rouge, LA 70803, USA}
\affiliation{Max Planck Institut f\"ur Gravitationsphysik, Albert
Einstein Institut, Am Muehlenberg 1, 14476 Golm, Germany}

\author{Ryoji Takahashi}
\affiliation{Center for Computation and Technology, 302 Johnston Hall,
Louisiana State University, Baton Rouge, LA 70803, USA}


\date{\today}


\begin{abstract}
  We perform both distorted black hole evolutions and binary black
  hole head on collisions and compare the results of using a full
  grid to results obtained by excising the black hole interiors. In
  both cases the evolutions are found to run essentially indefinitely,
  and produce the same, convergent waveforms. Further, since both the
  distorted black holes and the head-on collision of puncture initial
  data can be carried out without excision, they provide an excellent
  dynamical test-bed for excision codes. This provides a strong
  numerical demonstration of the validity of the excision idea, namely
  the event horizon can be made to ``protect'' the spacetime from the
  excision boundary and allow an accurate exterior evolution.
\end{abstract}


\pacs{
04.25.Dm, 
04.30.Db, 
95.30.Sf, 
97.60.Lf  
%
\\Preprint number: IGPG-04/10-5, AEI-2004-114
}


\maketitle


\section{Introduction}

A successful numerical model of a binary black hole collision will
require the refinement of a number of specialized techniques.  A
particular problem of black hole simulations is handling the strong
fields in the neighborhood of the physical singularity. The method of
``singularity excision'' promises to solve, or rather avoid, this
difficulty. Black hole excision was first discussed by
Thornburg~\cite{Thornburg85,Thornburg87,Thornburg91} based on a
suggestion by Unruh \cite{Unruh84}. Within the event horizon of a
spacetime all light-cones corresponding to the flow of physical
information in the fields point ``inwards''.  Thus a boundary within
the event horizon forms a one way membrane, with information flowing
only into the hole. Physical data on such a boundary should not
influence the external spacetime. In a numerical simulation, this
implies that errors on the boundary, for instance due to
extrapolations from the exterior, should not be expected to propagate
to the exterior spacetime.  However, because of the complicated and
nonlinear nature of the equations being solved, complex relationships
link the finite differenced evolution equations to artificial boundary
conditions. It is often the case that the full implications of using a
particular technique are poorly understood, and can only be judged by
some amount of experimentation.

The excision idea has a long history of implementations in numerical
relativity codes in 1D, 2D and 3D. The basic concept was developed
into a set of techniques for practical application in 1D
in~\cite{Seidel92a}, where causal differencing techniques and
dynamical lapse and shift conditions designed for excision were
introduced and shown to increase stability and accuracy of dynamic BH
spacetimes (coupled to self-gravitating scalar fields) by orders of
magnitude. This work was extended both to 3D BH
evolutions~\cite{Anninos94c} and to a large class of shift conditions
in~\cite{Anninos94e} such as the geometrically motivated distance
freezing, area freezing, expansion freezing and minimal distortion
shift. These ideas were later applied to more complex spacetimes,
e.g.\
\cite{Marsa96,Bruegmann96,Cook97a,Gomez98a,Walker98a,Brandt00,Shoemaker2003a},
typically using the Arnowitt-Deser-Misner (ADM) evolution system and
analytic shift conditions.

With the switch to the more stable BSSN evolution
system~\cite{Nakamura87,Shibata95,Baumgarte99,Alcubierre99d,Alcubierre99e}
and the accompanying improvements in stability of 3D codes, the
excision idea was revisited by Alcubierre and Br\"ugmann in the form
of an algorithm which they called ``Simple Excision''
\cite{Alcubierre00a,Alcubierre01a}. The idea was to attempt an
implementation which removed some complications that had hindered
previous efforts, namely causal differencing within the horizon, and
complicated interpolations onto an irregular boundary. Instead, the
excision region was fixed as a cubical region with faces along
constant Cartesian coordinate planes, simplifying the determination of
a normal, and a simple source-term-copy boundary condition replaced
true causal differencing. The result proved to be remarkably
successful at evolving distorted single black hole spacetimes allowing
accurate evolutions of unlimited length in time.

More recently advances in excision in 3D Cartesian coordinates have
been reported, using different techniques, for scalar fields in fixed
black hole backgrounds~\cite{Yo01b,Calabrese:2003a} (see also
\cite{Calabrese:2003vx}) and for evolutions of single black hole
spacetimes~\cite{Yo02a,Shoemaker2003a,
Sperhake-etal-2003-densitized-lapse-wobbling-BH-evolution}. In
general, the excision problem becomes much better controlled when
using coordinates that are adapted to the excision surface, and
furthermore when using hyperbolic formulations and making use of the
characteristic information about the propagation of all degrees of
freedom. Spherical excision is used in spectral
codes~\cite{Kidder00a,Kidder01a,Scheel2002a} but has also been
proposed for finite difference codes using multi-patch
techniques~\cite{Calabrese2003:excision-and-summation-by-parts,
Thornburg2004:multipatch-BH-excision,
Lehner-Reula-Tiglio-2004:multipatch-scalar-field-Kerr-background}.

A basic premise of the excision method is that the excision boundary
data should not propagate to the exterior spacetime. In practice, it
is important to verify that this is in fact the case for a given
numerical implementation, and to quantify any errors that might be
introduced. In this paper, we propose to compare long term black hole
evolutions using singularity excision to corresponding evolutions
carried out over a full grid, but with otherwise identical initial
data and evolution methods.  A useful excision method should yield
essentially identical evolutions, and thus identical physical
measurements, in the region of the spacetime outside of the horizon.

The main candidates for evolution without excision are methods using
singularity avoiding slicings, where the gauge condition on the lapse
slows evolution of the hypersurface in regions where a physical black
hole singularity is approached. In fact, singularity avoiding slicing
conditions (for example, maximal slicing or algebraic lapse conditions
of the ``1+log'' family) were the method of choice in most simulations
prior to the development of excision techniques. Without appropriate
shift conditions, slice stretching arises due to the differential
infall of coordinates, severely limiting the evolution time. It has
only very recently been shown that an appropriately chosen shift
vector can alleviate these problems, so that they occur at a much
later time (or even not at all for a fixed resolution) and the final
black hole settles to its expected stationary
state~\cite{Alcubierre02a,Reimann:2004yf}.

For this reason, comparisons between evolutions with and without
excision are rarely done.  Even worse, in most cases where excision
has been successfully used, very crude measures, such as ``time to
crash'', or some overall measure of the Hamiltonian constraint have
been used as a measure of success.  With the exception of a small
number of results~\cite{Walker98a,Alcubierre01a}, it has generally not
been verified that, for example, correct gravitational waveforms can be
extracted with excision, a point which is crucial for upcoming
gravitational wave observations.  As these waves will typically be
very small perturbations (of order $10^{-3}$ or less) of the 
metric~\cite{Camarda97b, Camarda97c,Allen98a}
very small errors generated by the excision techniques could easily
swamp the signals being extracted.  Furthermore, even if the excision
properly preserves the causal structure, allowing no physical signals
to propagate out from inside the horizon, gauge effects may well
propagate to the outside, influencing the solution.  This could show
up not only in metric functions but also in so-called
gauge-invariant waveforms, since such waveforms are only invariant
under {\em infinitesimal} gauge transformations given some assumptions
about the background coordinate system.  A large gauge wave
propagating through the spacetime may very well be seen in waves
extracted in this way.

In this paper, we propose to use both waveforms, extracted far from
the horizon, and detailed information extracted from apparent horizons
just outside the excision region, as important test quantities for
spacetimes evolved both with and without excision.  Such tests would
have been difficult or
impossible in the past because, until recently, 3D BH simulations
without excision became rapidly inaccurate and crashed after evolution
times of only $t\approx 30-40M_{ADM}$, where $M_{ADM}$ is the ADM
mass.  However, the recent development of powerful shift conditions
has to some extent cured the problem of slice stretching associated
with singularity avoiding slicing, making it possible to evolve
certain classes of distorted and colliding BH spacetimes much longer
(for thousands of $M_{ADM}$) and much more accurately than ever before,
without the need for excision~\cite{Alcubierre02a}.  This makes it
possible for the first time to carry out systematic, long term studies
of the effects of the excision technique on the evolved BH spacetimes.

Following Refs.~\cite{Camarda97b,Camarda97c,Allen97a,Allen98a} we use
distorted black hole initial data sets, whose waveforms can be
independently and reliably computed, as testbeds for the excision
techniques we use.  Those papers showed that even modes with energies
of order $10^{-7}M_{ADM}$ could be very accurately extracted in a full
3D numerical evolution, as confirmed by comparisons with purely
perturbative evolutions.  We will use some of the same testbeds
proposed there in this paper.

We also apply the ideas to the study of BH collisions, using
Brill-Lindquist-type initial data which, as showed in
Ref.~\cite{Alcubierre02a}, can be evolved indefinitely without
resorting to excision. Although Misner data have been much more
extensively studied~\cite{Anninos93b,Anninos94b,Price94b} and would be
an excellent test-bed, we use Brill-Lindquist here because of their
close relation to the Brill-Lindquist family of ``puncture'' data
often applied in generating data for binary BHs in quasi-circular
orbit~\cite{Brandt97b,Baumgarte00a,Tichy02,Tichy03a,Tichy03b}. 
These data sets possess apparent horizons in the initial slice,
allowing excision regions to be defined, but have also been evolved
for unlimited times without excision (after merger). As such, these
evolutions provide an ideal test-bed for excision methods. If the
excision method is valid and the exterior spacetime is protected from
errors on the excision boundary by the event horizon, then simulations
with excision should accurately reproduce the non-excised results.
Because the testbeds we use here are highly dynamic, distorted and
colliding black holes, evolved in general coordinate systems in 3D
Cartesian coordinates, without resort to special conditions or tricks,
successful tests of excision techniques against independently computed
physics results would provide a certain amount of confidence that the
techniques can be applied to more complex systems, such as orbiting
black holes, where the detailed physics is more critical to get right.

In the next section we briefly outline the methods used in our black
hole simulations. There follows a description of the excision region
and boundary conditions which we are applying. Finally, we present
numerical results in which distorted and binary black hole evolutions
using excision are compared directly with those performed without
excision.


\section{Model and Methods}  

\subsection{Initial Data and Evolution Methods}

The details of the formulation which we have implemented numerically
are outlined in~\cite{Alcubierre02a}. Briefly, we make use of the
``BSSN'' evolution system~\cite{Nakamura87, Shibata95, Baumgarte99}.
The evolution variables are a conformal factor, the conformally
decomposed metric, the trace-free extrinsic curvature tensor and its
trace. Additionally, the contracted conformal Christoffel symbols
$\tilde\Gamma^i:=\tilde \gamma^{jk} \tilde \Gamma^i_{jk}$ are evolved
as independent variables. These variables can be regarded as gauge
source functions, and in our evolutions are used to fix the shift
vector through e.g. the condition $\partial_t \tilde\Gamma^i=0$, which
results in an elliptic shift condition analogous to minimal
distortion~\cite{Smarr78b}. In practice, instead of applying this
elliptic shift condition directly, we use a hyperbolic ``Gamma
driver'' version as described in Eq. (46) of~\cite{Alcubierre02a},
\begin{equation}
\partial^2_t \beta^i = F \, \partial_t \tilde\Gamma^i - 
(\eta - \frac{\partial_t F}{F}) \, \partial_t \beta^i.
\label{eq:gamma0}
\end{equation}
The parameter $\eta$ is a driver term which controls the growth of the
shift and is discussed in more detail later, and $F$ is a function
which can be used to condition its overall shape.

The black hole initial data we use comes in two classes.  The first
class is the ``Brill wave plus black hole'' family of distorted BHs,
constructed topologically as wormholes isometrically connecting two
asymptotically flat spacetimes.  These data sets have proved a rich
and powerful system for developing many techniques in numerical
relativity and for studying the physics of many aspects of distorted
BHs~\cite{Bernstein94a,Abrahams92a, Anninos93a, Bernstein93b,
Brandt94c,Brandt94a,Anninos94f,Camarda97b, Camarda97c, Allen98a,
Alcubierre01a,Brandt02a}.

The second class is Brill-Lindquist binary BH initial data in
isotropic coordinates~\cite{Brill63}. The topology of these data sets
is that of an asymptotically flat spacetime connected through
wormholes to two disconnected asymptotically flat ends. These data sets
are time symmetric and set the stage for BH head-on collisions.

As discussed in~\cite{Bruegmann97,Alcubierre02a}, for spacetimes in
which there is enough symmetry (such as the axisymmetric head-on
collision), it is possible to carry out expansions of the evolution of
the metric quantities at the ``punctures'' (the points in ${\cal R}^3$
at the ``center'' of each BH representing the compactified
asymptotically flat regions on the other side of the wormholes), thus
ensuring that the evolution variables remain regular there. By careful
consideration of the evolution at the punctures, it is therefore
possible to carry out long-term evolutions with the punctures included
on the grid, i.e.\ without excision.


\subsection{Lego Excision}  The excision boundary condition which we apply
is a variation of the ``Simple Excision'' methods described
in~\cite{Alcubierre00a}. The important difference is that instead
of excising a cubical region, we excise an irregular region. The size
and shape of the excision region is determined dynamically by the
location of the apparent horizon. On a Cartesian grid, the resulting
surface is a blocky quasi-spherical region, prompting the name ``lego
excision'' for this method.

The update of all interior points on the grid (including those within
the apparent horizon) is carried out using centered differences. That
is, no causal differencing is applied. The exceptions are advection
terms on the shift (terms of the form $\beta^i \partial_i \,\,$), for
which a second order upwind in the shift direction is applied.

For points on the excision boundary itself, a centered update scheme
cannot be used for lack of data to one side. Instead, data at these
points are updated using a time derivative term which is copied from
one point out. This has the effect of allowing a certain amount of
evolution to take place on the boundary, while also allowing the
boundary to ``settle down'' as the system settles to what should be a
stationary end state, exemplified by $\partial_t\phi = 0$ for the
fields $\phi$~\footnote{Of course, with a poor choice of coordinates
and gauge it may be the case that the stationarity takes on a very
different form, however in practice we have found that the gauge
choices which we apply do drive the system to a state in which the
time derivatives tend to zero~\cite{Alcubierre02a}}. The neighboring
point from which the time derivative is copied is determined by a
normal to the surface at the point. The time derivative from the point
nearest to the normal is copied to the boundary point.

For the models in question, we assume that the excision region neither
shrinks nor moves across the grid. In other words, it is not possible
for grid points to emerge from the excision region: once excised, a
point remains excised. While this may at first seem like a strong
limitation, in fact even with these restrictions the technique is 
adequate for many binary BH inspiral studies. Under reasonable gauge choices it
should be expected that the BH horizons individually should either
remain of an approximately fixed coordinate size, or grow slightly, as
is the case for the final post-merger BH.  Further, for binary BH
inspirals, co-rotating coordinates reduce the amount of dynamics on
the grid to the dynamical time-scale of the problem in question, and
either remove or greatly reduce the orbital motion of the individual
bodies.  Co-rotating coordinate systems have been proposed (see, e.g.~\cite{Thorne98a}), and
implemented for cases of BH
binaries~\cite{Alcubierre2003:co-rotating-shift,Alcubierre2003:pre-ISCO-coalescence-times,Bruegmann:2003aw}
(for neutron stars see e.g.\ \cite{Duez:2002bn}).  For the puncture
data which we have used here, the punctures are fixed to
the grid. 
As such, problems involving movement of the excision region do not arise in these cases, almost by
construction.


\section{Applications}

\subsection{Results for Distorted Black Hole}

As a first application we study a simulation of a distorted BH which
has been studied extensively in previous
work~\cite{Camarda97b,Camarda97c,Allen97a,Allen98a,Bernstein94a,Abrahams92a,Alcubierre01a}.
The BH is initially distorted by an even parity Brill wave and, for a 
low enough amplitude wave, during
the evolution the BH rings at quasi-normal frequencies before settling down
to a Schwarzschild BH. The parameters used here are, in the notation
of~\cite{Brandt97c,Bernstein94a,Abrahams92a,Alcubierre01a}, $Q_0=0.1$,
$\eta_0=0$, $\sigma=1$. For this simulation the computational domain
extends to $\pm 34.80$ (in coordinate units).
For these initial data parameters, $M_{ADM}=1.92$, which puts the outer
boundary at $\pm 18.13M_{ADM}$. We also use two different resolutions,
the smallest one with a $288\times 288\times 144$ sized uniform grid
and a coordinate resolution of $0.24$, and the largest with a $384\times
384\times 192$ sized uniform grid and a resolution of $0.18$. An
explicit reflection symmetry about the $z=0$ plane is used.

We evolve this system both with and without excision, using identical
evolution parameters.  For the gauge we use a Gamma driver shift condition as specified in
Eq.~(\ref{eq:gamma0}), and 1+log slicing. In the simulation using excision, the lego
excision region was located at 80\% of the apparent horizon radius, with a
minimum buffer size of at least 5 grid points between the excision
region and the apparent horizon surface. 
We find the horizon using methods
described in~\cite{Schnetter02a,Thornburg2003:AH-finding}, and implemented as 
in~\cite{Thornburg2003:AH-finding}.
At higher resolution the size
of the buffer was adjusted such that the excision region remained at
approximately the same position.

\begin{figure}
\epsfxsize=80mm
\epsfbox{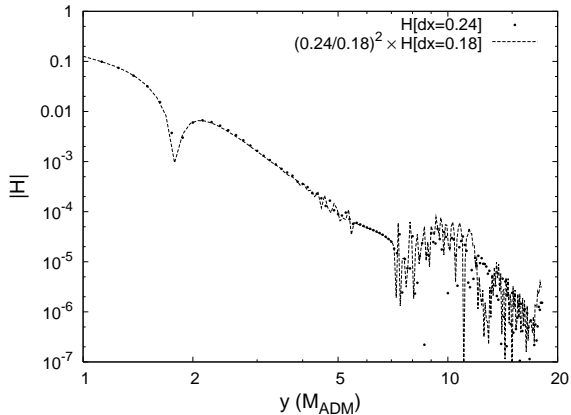}
\caption{The absolute value of the Hamiltonian constraint
$|H|$ for a distorted black hole simulation without excision,
evaluated along the $y$-axis at $T=9.38M_{ADM}$
($x=z=0$). The high resolution Hamiltonian is scaled such that the two
curves should coincide for second order convergence.}
\label{fig:conv-no-exc}
\end{figure}

\begin{figure}
\epsfxsize=80mm
\epsfbox{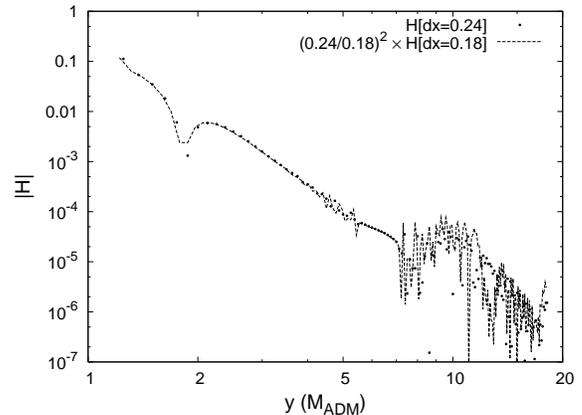}
\caption{Absolute value of the Hamiltonian constraint $|H|$ along the
$y$-axis for the same distorted black hole simulation shown in
Fig.~\ref{fig:conv-no-exc}, but this time using excision. The
Hamiltonian constraint was computed at $T=9.38M_{ADM}$.}
\label{fig:conv-exc}
\end{figure}

In Fig.~\ref{fig:conv-no-exc} and Fig.~\ref{fig:conv-exc} we show
convergence plots for the absolute value of the Hamiltonian constraint
along the $y$-axis at time $T=9.38M_{ADM}$, for the runs without and
with excision, respectively. In both plots the higher resolution case
is scaled by a factor of $(0.24/0.18)^2$ so that, in case of second
order convergence, the two curves should coincide. Note that, in both
cases, we have second order convergence in the regions that
are causally disconnected from the outer boundaries. In
Fig.~\ref{fig:conv-exc} points inside the excision region have been
removed from the plot. Note that outside the excision region the curve
is very similar to that of Fig.~\ref{fig:conv-no-exc}. The noise
visible in both plots at $y>7M_{ADM}$ is due to numerical details of
the initial data constraint solver.

\begin{figure}
\epsfxsize=80mm
\epsfbox{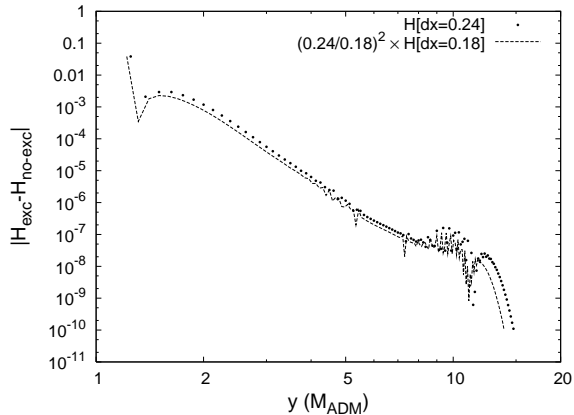}
\caption{Differences in the Hamiltonian constraints between the runs
with and without excision $|H_{\rm exc}-H_{\rm no-exc}|$ at
$T=9.38M_{ADM}$ for the distorted black hole. The difference between the
two cases converges away at second order.}
\label{fig:exc-against-no-exc}
\end{figure}

In Fig.~\ref{fig:exc-against-no-exc} we plot the difference between
the Hamiltonian constraint between runs with and without excision at
$T=9.38M_{ADM}$ at the two different resolutions, with the high
resolution case again scaled as in the previous plots. Only the points
outside the excision region are shown in the plot. The figure shows
that there are indeed small differences between the two cases, but
that these differences converge away at second order outside of
the excision region.

\begin{figure}
\epsfxsize=80mm \epsfbox{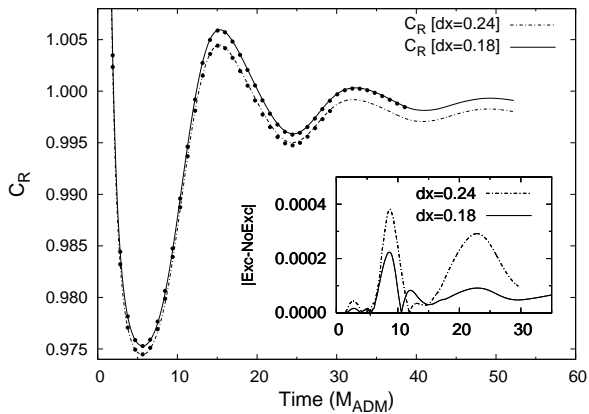}
\caption{Ratio $C_r$ of the polar and equatorial circumference of the
apparent horizon for two different resolutions of the distorted black
hole simulation. The lines denote the excision data and the dots show
the results of runs without excision. The inset shows the absolute
value of the difference between the two cases.}
\label{fig:distorted-bh-cr}
\end{figure}

As an indicator for analyzing excision effects we use
the ratio $C_r$ of the circumference along a polar direction in the {\em xz}-plane,
to the circumference in the equatorial {\em xy}-plane, of the apparent
horizon, as a function of time. As these measurements are made very close to the excision
region, and as for the low amplitude Brill wave studied here $C_r$ remains
within about 1\% of the Schwarzschild value of unity, we regard this as
a very sensitive indicator for these tests.  
In Fig.~\ref{fig:distorted-bh-cr}, these ratios are plotted
for evolutions with excision (lines) and without (dots) at two
resolutions. The differences are very small in comparison to the size
of the physical wave, and much smaller with increased of resolution.
Note that in this plot the non-excision curves do not extend over the
same time as the excision data. The earlier
termination of the non-excision runs in Fig.~\ref{fig:distorted-bh-cr} was 
due to a hardware problem and was not caused by problems in the code.

\begin{figure}
\epsfxsize=80mm
\epsfbox{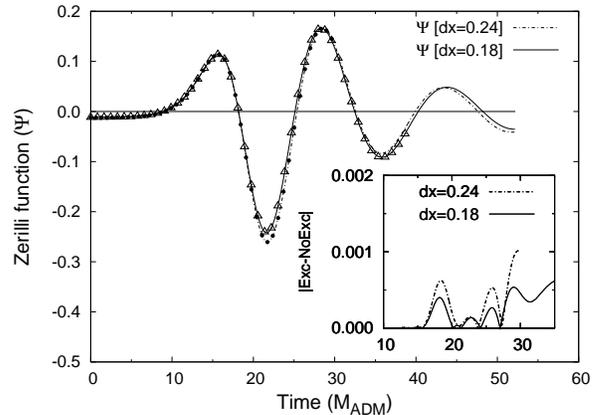}
\caption{Zerilli function $\psi$ ($\ell=2$,$m=0$ mode) calculated at
\mbox{$r=13.5M_{ADM}$}. The lines show the excision data and the points
show the results of the simulations without excision. The inset shows
the absolute value of the difference in the waveforms which decreases
as the resolution is increased.}
\label{fig:distorted-bh-zerilli-exc-no-exc}
\end{figure}

In Fig.~\ref{fig:distorted-bh-zerilli-exc-no-exc} we show the
evolution of the Zerilli wave function $\psi$ ($\ell=2$, $m=0$
mode). The figure shows excision data as lines and the results from
simulations without excision as points. The absolute value of the
difference between the waveforms is also shown in the inset. This
difference is much smaller than the Zerilli function itself and decreases
with resolution.

\begin{figure}
\epsfxsize=80mm
\epsfbox{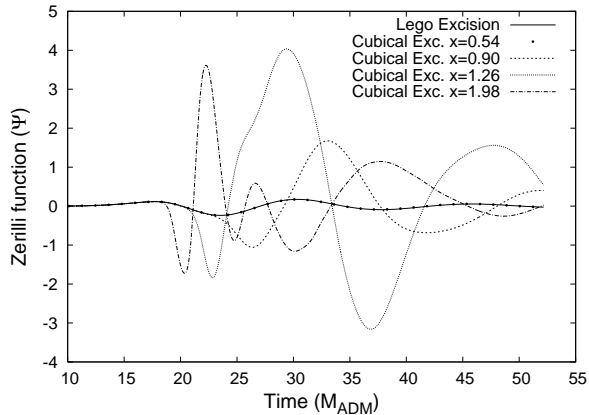}
\caption{Zerilli Function $\psi$ for the distorted black hole
calculated at $r=15.3M_{ADM}$, using a resolution $dx=0.18$ and
different cubical excision box sizes. For the first two evolutions,
the excised region is within the horizon and show essentially
identical results for both the cubical and ``lego'' excision
region. The other evolutions have at least a portion of the excision
box extending outside the horizon, and thus do not represent true
outflow boundaries. They show dramatic differences from the previous
waveforms.}
\label{fig:distorted-bh-exc-box}
\end{figure}

We also studied the effects of the size of the excision box for this
model. One can expect that problems will occur if the excision box is
made too large and in particular if it reaches outside of the black
hole. In Fig.~\ref{fig:distorted-bh-exc-box} we show the influence of
differently sized excision boxes on the waveform. For these
simulations the black hole is nearly spherical (see
Fig.~\ref{fig:distorted-bh-cr}) with a mean coordinate radius of
initially about $1$ which then grows to about $2$ during the first
$5M_{ADM}$ of evolution. For comparison we have also used a cubical
excision box which remained fixed throughout the evolution. The
resolution used in these simulations was $0.18$. We first plot the
waveform of the lego excision simulation used above. Next we show
waveforms for cubical excision regions where half the length
of one side of the cube is $0.54$, $0.90$, $1.26$ and $1.98$,
respectively. For the case with the smallest excision box the waveform
remains essentially unaffected.  However, once the excision region
becomes larger and even extends outside of the initial apparent horizon a
large effect on the waveform becomes visible. This demonstrates that
excision must be done inside the horizon as one would expect. It also
indicates that with techniques implemented here, a buffer zone of two to three points in the initial hole
can be sufficient to accurately extract gravitational waves.

\subsection{Results for Head-on Collision}

As a second example we consider a binary BH head-on collision for
Brill-Lindquist type data, starting from a close coordinate separation
between punctures of $2.303$. For this data set each individual BH
has a mass parameter of $M_i=0.5$ for a total ADM mass of 
$M_{ADM} = 1$~\cite{Brill63}. The
initial data uses a ``fish-eye'' coordinate transformation to push the
boundaries further out~\cite{Baker00b,Alcubierre02a}. We have
performed runs using octant symmetry with central resolutions of
$0.128$, $0.064$, and $0.032$, on cubical grids with $96^3$,
$192^3$ and $384^3$ grid points respectively. Using the ``fish-eye''
parameters $a=3$, $s=1.2$ and $r_0=5.5$, in the notation
of~\cite{Alcubierre02a}, places the coordinate boundaries at $12.288$
and the physical boundaries at $25.862$.

The evolution without excision is identical to the one shown in
Figs.~11 and 12 of Ref.~\cite{Alcubierre02a}. However, running
the excision case with exactly the same parameters turned out to be
impossible. The reason is that with the choice of damping coefficient
$\eta$ used in that reference for the Gamma driver shift condition,
the lapse and shift turned out to be too dynamic at the excision
boundary, causing the simulation to crash very early.  We have found
that this effect can be controlled by using a smooth spatially varying
$\eta$-parameter.  This was implemented using the
conformal factor from the initial data
\begin{equation}
\Psi = 1 + \frac{M_1}{2r_1} + \frac{M_2}{2r_2},
\end{equation}
with $M_1$ and $M_2$ the mass parameters of the black holes (in this
case $M_1=M_2=0.5$) and $r_1$ and $r_2$ the coordinate distances to
the two punctures. A smoothly varying $\eta$-parameter can then be
constructed (this construction is of course not unique) as
\begin{equation}
\eta = \eta_{\mathrm{punc}}-\frac{\eta_{\mathrm{punc}} -
       \eta_{\infty}}{C \left(1+(\Psi-1)^2\right )},
\end{equation}
where $\eta_{\mathrm{punc}}$ is the value of $\eta$ near the
punctures, $\eta_{\infty}$ is the value of $\eta$ at infinity, and $C$
is a parameter that controls the width of the transition region.  Here
we used $\eta_{\mathrm{punc}}=5.6$, $\eta_{\infty}=2.8$ and
$C=1$.

\begin{figure}
\epsfxsize=80mm
\epsfbox{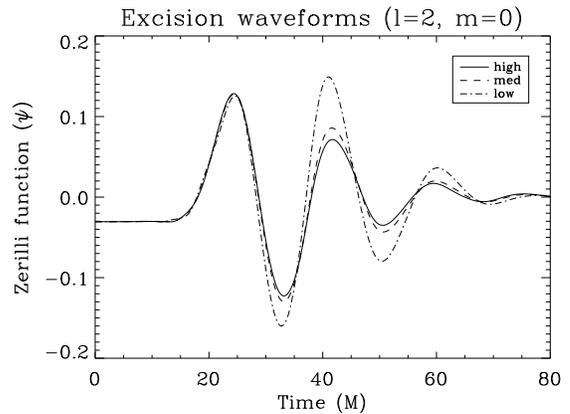}
\caption{Zerilli function $\psi$ ($\ell=2$, $m=0$ mode) calculated at
\mbox{$r=14.8M_{ADM}$} for the head-on collision using excision at
resolutions 0.128 (low), 0.064 (med) and 0.032 (high).}
\label{fig:bbh0_waves}
\end{figure}

\begin{figure}
\epsfxsize=80mm
\epsfbox{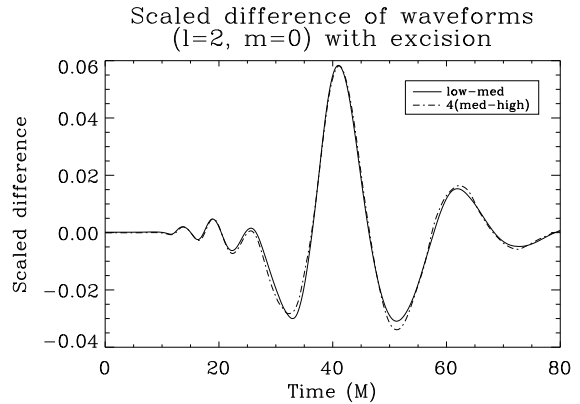}
\caption{Scaled difference of the Zerilli function $\psi$
($\ell=2$, $m=0$ mode) calculated at \mbox{$r=14.8M_{ADM}$} for the head-on
collision using excision showing second order convergence.}
\label{fig:convergence}
\end{figure}

The resulting waveforms for the excision runs, using the above
construction for $\eta$ and the three different resolutions, can be
seen in Fig.~\ref{fig:bbh0_waves}.  These waveforms are second order
convergent as can be seen from Fig.~\ref{fig:convergence}, where the
difference between the low and medium resolution waveforms and four times
the difference between the medium and high resolution waveforms are
plotted together.

\begin{figure}
\epsfxsize=80mm
\epsfbox{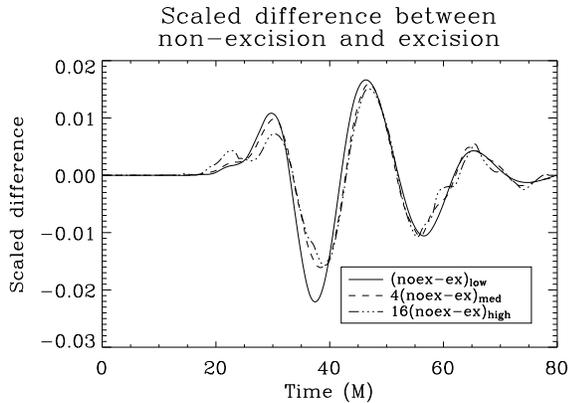}
\caption{Scaled difference in waveforms from evolutions with
and without excision for the three different resolutions used in
the head-on collision.}
\label{fig:match}
\end{figure}

At all three resolutions we can see some differences in the waveforms
between the runs with and without excisions. There are basically two
reasons for this.  The first is of course the presence of the excision
boundary and the second is the slightly different gauge choices that
were needed in each case in order to obtain long enough
evolutions. However, as can be seen from Fig.~\ref{fig:match}, these
differences converge away to second order with resolution (except for
a small gauge effect which is visible in the initial part of the
waveform).

For the above runs, we used an excision region which was 80\% of the
apparent horizon radius with minimum buffer size of (5,10,20)
grid points for the three different resolutions. This was in order to
ensure that we had a buffer zone large enough suppress inaccuracies at
the excision boundary from propagating out of the hole, as well as to
ensure that the apparent horizon can always be located, as this
requires points on both sides of the surface.

We note that the runs performed with excision are more sensitive to
gauge parameters than the evolutions without excision. This is essentially due to
the fact that the shift can be used to control the rate of growth of
the horizon, and in some cases can even draw it inwards. For the
excised runs, this means that the horizon can be brought too close to
the excision boundary, allowing errors to ``leak out''. The shift
parameters used in this paper allow a slightly faster growth of the
horizon than those used in~\cite{Alcubierre02a}.


\section{Conclusions} 

We have proposed a sensitive test-bed for 3D black hole excision
techniques in numerical relativity, whereby simulations with and
without excision are carried out on identical initial data sets.  A
recently developed family of powerful gauge conditions now allows long
evolutions of some black hole spacetimes without excision, making such
a test-bed practical for the first time.  Applying these ideas to our
own methods, we demonstrated that our excision procedure is not only
effective in improving the length of time that black holes can be
evolved, but also that it preserves sensitive details of the physics
that can be extracted. We carried out both distorted and binary black
hole head-on collision simulations, both with and without excision. We
used both the ratio of apparent horizon circumferences and the very
sensitive indicator of extracted waveforms to compare physically
important details of the simulations. In particular, the waveforms
represent very small signals buried in the metric functions. The
results are essentially identical, indicating that the excision
boundary condition has preserved the accuracy of the
runs.

The evolutions were carried out for cases of
axisymmetric initial data, where for singularity avoiding slicing with
an appropriate shift condition the evolution variables could be
suitably controlled at the locations of the punctures. For more
general situations it can be more difficult to maintain a regular
evolution. In such situations, even for the puncture data discussed
above, it is expected that excision would be necessary for a
long-term evolution.  However, the tests shown here have given us
confidence that the excision methods which we are using are robust for
reasonably complicated situations, and do not adversely affect the
accuracy of evolutions.

Runs without excision form an important benchmark against which the
effects of excision techniques can be analyzed. Evolutions of black
hole spacetimes involve a patchwork of experimental techniques, all of
which are needed to obtain long evolution times, but each of which
carries with it complications that can affect the accuracy and
stability of the simulation, both in isolation and in interaction with
the entire system. Given this fact, the models considered above can
form a particularly important test of the effects of applying an
excision boundary condition.

The initial data sets we used here are readily available to anyone
through the Cactus framework. This will make it possible for other
groups to apply the same test-bed to their own evolution codes.


\begin{acknowledgments}

The lego excision code discussed in this paper was developed in
collaboration with Erik Schnetter and Deirdre Shoemaker.  Results for
this paper were obtained using computing time allocations at the AEI,
CCT, LRZ, NCSA, NERSC, PSC and RZG. We use Cactus and the 
\texttt{CactusEinstein}
infrastructure with a number of locally developed thorns.  This work
was supported in part by NSF grants PHY-02-18750 and PHY-02-44788, by
the DFG grant ``SFB/Transregio~7: Gravitational Wave Astronomy'',
by DGAPA-UNAM grants IN112401 and IN122002, and by the EU Programme
`Improving the Human Research Potential and the Socio-Economic Knowledge
Base' (Research Training Network Contract HPRN-CT-2000-00137).

\end{acknowledgments}


\bibliographystyle{apsrev}

\bibliography{references}


\end{document}